\documentclass{PoS1}

\title{Numerical approach to multi-loop integrals}

\ShortTitle{Numerical approach to multi-loop integrals}

\author{\speaker{K. Kato}$^a$\thanks{E-mail:{\tt kato@cc.kogakuin.ac.jp}}, 
E.~de~Doncker$^b$, 
N.~Hamaguchi$^c$, 
T.~Ishikawa$^c$, 
T.~Koike$^d$, 
Y. Kurihara$^c$, 
Y.~Shimizu$^c$, 
and F.~Yuasa$^c$ \\
\llap{$^a$} Kogakuin University, 1-24 Nishi-Shinjuku, Shinjuku, Tokyo 163-8677, Japan \\
\llap{$^b$} Western Michigan University Kalamazoo, MI 49008-5371, USA \\
\llap{$^c$} High Energy Accelerator Research Organization (KEK), 1-1 Oho Tsukuba, Ibaraki 305-0801, Japan\\
\llap{$^d$} Seikei University, Musashino, Tokyo 180-8633, Japan \\
}

\abstract{For the calculation of multi-loop Feynman integrals, 
a novel numerical method, the Direct Computation Method (DCM) is 
developed. 
It is a combination of a numerical integration and a
series extrapolation. In principle, DCM can handle diagrams of
arbitrary internal masses and external momenta, and can
calculate integrals with general numerator function.
As an example of the performance of DCM, a numerical computation of
two-loop box diagrams is presented. 
Further discussion is given on the choice of
control parameters in DCM.  This method will be an indispensable tool
for the higher order radiative correction when it is
tested for a wider class of physical parameters and the separation
of divergence is done automatically.}

\FullConference{%
The XXth International Workshop High Energy Physics and Quantum Field Theory \\
September 24-October 1, 2011\\
Sochi Russia}

\newcommand{\siki}[1]{Eq.\ref{eq:#1}}
\newcommand{\zu}[1]{Fig.\ref{fig:#1}}

\newif\iffigureon
\figureontrue

\begin{document}

\section{Introduction}
\label{sec:intro}

The high-statistics data 
in high-energy physics requires the
theoretical prediction with enough accuracy.
The prediction can be given by 
perturbative calculation in quantum field theory.
Then the multi-loop integral is an indispensable 
component for the theoretical study.

We define the multi-loop integrals by the
following formula where the space-time dimension
is denoted as $n=4-2\delta$\ \footnote{
The symbol $\epsilon$ is reserved for the (infinitesimal) parameter
in the propagator.
}.
Here we confine the discussion to scalar integrals only.
Since the method presented below is basically
numerical, the inclusion of a numerator will be straight-forward.
\begin{equation}
{\cal I}=\int \prod_{j=1}^L \frac{d^n \ell_{j}}{(2\pi)^n i} \ 
  \prod_{r=1}^N\frac{1}{D_r}
\end{equation}
where the propagator is $D_r=q_r^2-m_r^2 + i\epsilon$, 
$N$ is the number of propagators and
$L$ is the number of loops.
We combine the propagators by the standard Feynman
parameter integral
\begin{equation}
\prod_{r=1}^N\frac{1}{D_r}=(N-1)! \int \prod dx_{r} 
\frac{\delta(1-\sum x_r)}{(\sum x_r D_r)^N}
\end{equation}
and perform integration with respect to the loop momenta.
We obtain
\begin{equation}
{\cal I}=\frac{\Gamma(N-nL/2)}{(4\pi)^{nL/2}}\times I,\quad
I=(-1)^N
\int \prod dx_{r} 
\frac{\delta(1-\sum x_r)}{U^{n/2}(V-i\epsilon)^{N-nL/2}}\,,
\label{eq:integral}
\end{equation}
\begin{equation}
V=M^2-\frac{W}{U},\qquad M^2= \sum_r x_r m_r^2
\end{equation}
where $U$ and $W$ are polynomials in the $x$ parameters\cite{basic}.

In Section 2, we propose a unique method to calculate the
integral $I$ in \siki{integral}. 
We call the method the Direct Computation Method (DCM).
In the preceding works\cite{works}, DCM has successfully calculated
one-loop and two-loop diagrams.
As an example we show the results for the two-loop box diagrams
in Section 3 and also report a study on the parameters in DCM.
We discuss further aspects of DCM in Section 4.


\section{Method}
\label{sec:method}

DCM consists of a regularized numerical integration
and an extrapolation of a numerical sequence.

\begin{figure}[htb]
\begin{center}
\unitlength 0.1in
\begin{picture}( 48.0000, 13.9000)(  4.0000,-14.2000)
%
\special{pn 8}%
\special{pa 400 800}%
\special{pa 2400 800}%
\special{fp}%
\special{sh 1}%
\special{pa 2400 800}%
\special{pa 2334 780}%
\special{pa 2348 800}%
\special{pa 2334 820}%
\special{pa 2400 800}%
\special{fp}%
%
\special{pn 8}%
\special{pa 3200 800}%
\special{pa 5200 800}%
\special{fp}%
\special{sh 1}%
\special{pa 5200 800}%
\special{pa 5134 780}%
\special{pa 5148 800}%
\special{pa 5134 820}%
\special{pa 5200 800}%
\special{fp}%
%
\special{pn 8}%
\special{pa 600 800}%
\special{pa 600 800}%
\special{fp}%
\special{pa 600 750}%
\special{pa 600 750}%
\special{fp}%
%
\special{pn 8}%
\special{pa 600 800}%
\special{pa 600 730}%
\special{fp}%
%
\special{pn 8}%
\special{pa 2200 800}%
\special{pa 2200 800}%
\special{fp}%
\special{pa 2200 750}%
\special{pa 2200 750}%
\special{fp}%
%
\special{pn 8}%
\special{pa 2200 730}%
\special{pa 2200 730}%
\special{fp}%
\special{pa 2200 780}%
\special{pa 2200 780}%
\special{fp}%
%
\special{pn 8}%
\special{pa 2200 730}%
\special{pa 2200 800}%
\special{fp}%
%
\special{pn 8}%
\special{pa 3400 800}%
\special{pa 3400 800}%
\special{fp}%
\special{pa 3400 750}%
\special{pa 3400 750}%
\special{fp}%
%
\special{pn 8}%
\special{pa 3400 800}%
\special{pa 3400 730}%
\special{fp}%
%
\special{pn 8}%
\special{pa 5020 970}%
\special{pa 5020 970}%
\special{fp}%
\special{pa 5020 920}%
\special{pa 5020 920}%
\special{fp}%
%
\special{pn 8}%
\special{pa 5000 800}%
\special{pa 5000 730}%
\special{fp}%
%
\special{pn 20}%
\special{pa 970 770}%
\special{pa 1020 820}%
\special{fp}%
\special{pa 1020 770}%
\special{pa 970 820}%
\special{fp}%
%
\special{pn 20}%
\special{pa 1770 780}%
\special{pa 1820 830}%
\special{fp}%
\special{pa 1820 780}%
\special{pa 1770 830}%
\special{fp}%
%
\special{pn 20}%
\special{pa 4580 180}%
\special{pa 4630 230}%
\special{fp}%
\special{pa 4630 180}%
\special{pa 4580 230}%
\special{fp}%
%
\special{pn 20}%
\special{pa 4580 380}%
\special{pa 4630 430}%
\special{fp}%
\special{pa 4630 380}%
\special{pa 4580 430}%
\special{fp}%
%
\special{pn 20}%
\special{pa 4580 580}%
\special{pa 4630 630}%
\special{fp}%
\special{pa 4630 580}%
\special{pa 4580 630}%
\special{fp}%
%
\special{pn 20}%
\special{ar 1000 800 90 90  3.1415927 6.2831853}%
%
\special{pn 20}%
\special{ar 1800 790 100 100  0.0665682 3.1415927}%
%
\special{pn 20}%
\special{pa 600 790}%
\special{pa 900 790}%
\special{fp}%
\special{pa 2200 790}%
\special{pa 1910 790}%
\special{fp}%
%
\special{pn 20}%
\special{pa 1100 790}%
\special{pa 1700 790}%
\special{fp}%
%
\special{pn 20}%
\special{pa 3400 800}%
\special{pa 5000 800}%
\special{fp}%
%
\special{pn 20}%
\special{pa 3780 970}%
\special{pa 3830 1020}%
\special{fp}%
\special{pa 3830 970}%
\special{pa 3780 1020}%
\special{fp}%
%
\special{pn 20}%
\special{pa 3780 1170}%
\special{pa 3830 1220}%
\special{fp}%
\special{pa 3830 1170}%
\special{pa 3780 1220}%
\special{fp}%
%
\special{pn 20}%
\special{pa 3780 1370}%
\special{pa 3830 1420}%
\special{fp}%
\special{pa 3830 1370}%
\special{pa 3780 1420}%
\special{fp}%
%
\special{pn 8}%
\special{pa 4490 190}%
\special{pa 4490 590}%
\special{fp}%
\special{sh 1}%
\special{pa 4490 590}%
\special{pa 4510 524}%
\special{pa 4490 538}%
\special{pa 4470 524}%
\special{pa 4490 590}%
\special{fp}%
%
\special{pn 8}%
\special{pa 3910 1400}%
\special{pa 3910 1000}%
\special{fp}%
\special{sh 1}%
\special{pa 3910 1000}%
\special{pa 3890 1068}%
\special{pa 3910 1054}%
\special{pa 3930 1068}%
\special{pa 3910 1000}%
\special{fp}%
\put(6.0000,-10.0000){\makebox(0,0)[lb]{0}}%
\put(34.0000,-10.0000){\makebox(0,0)[lb]{0}}%
\put(22.0000,-10.0000){\makebox(0,0)[lb]{1}}%
\put(50.0000,-10.0000){\makebox(0,0)[lb]{1}}%
\put(23.4000,-7.3000){\makebox(0,0)[lb]{$x$}}%
\put(51.5000,-7.3000){\makebox(0,0)[lb]{$x$}}%
\put(48.0000,-2.0000){\makebox(0,0)[lb]{$x_1+i\varepsilon_0$}}%
\put(48.0000,-4.0000){\makebox(0,0)[lb]{$x_1+i\varepsilon_1$}}%
\put(48.0000,-6.0000){\makebox(0,0)[lb]{$x_1+i\varepsilon_2$}}%
\put(4.1000,-4.1000){\makebox(0,0)[lb]{(a)}}%
\put(32.0000,-4.0000){\makebox(0,0)[lb]{(b)}}%
\end{picture}%
\end{center}
\caption{Integral paths for (a) analytical calculation and for (b) DCM.
The cross ($\times$) stands for the singularity of integrand. }
\label{fig:simple}
\end{figure}
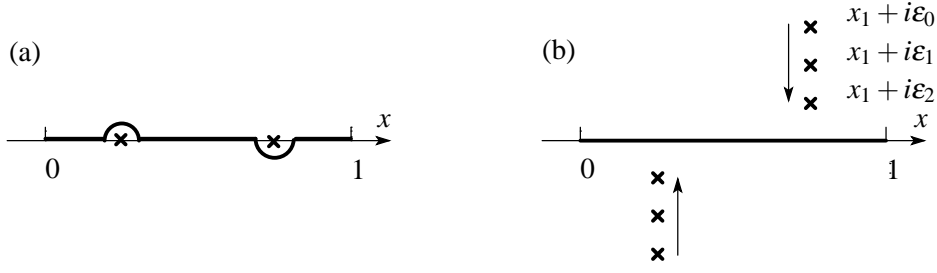

In order to illustrate the idea of regularized
numerical integration, let us consider a simple integral:
\begin{equation}
J=\int_0^1 \frac{dx}{m^2-sx(1-x)-i\epsilon}
\label{eq:simple}
\end{equation}

As is shown in \zu{simple}, the integrand of \siki{simple} has
singular points when $s>4m^2$. Analytically, this can be
handled by taking $\epsilon$ as an infinitesimal positive
quantity, or, in other words, by the hyper-function formula 
$1/(z-i\epsilon)=P(1/z)+i\pi\delta(z)$.
Numerical computation is unstable if the integrand is
divergent at some points. A way to solve the situation
is to deform the path in the complex $x$ plane to avoid
the singular points. However, the deformation would be
very complicated for the integrand of a multi-variable function.
Another possibility is to assume that $\epsilon$ is a finite quantity
as is shown in 
\zu{simple}(b).  Then, the numerical integration along the
$x$-axis is stable. If we can calculate the limiting value
for $\epsilon\rightarrow 0$, it is the value of the integral $J$.

The loop integral $I(\epsilon)$ is defined by
\begin{equation}
I(\epsilon)=(-1)^N
\int \prod dx_{r} 
\frac{\delta(1-\sum x_r)(V+i\epsilon)^{N-nL/2}}%
{U^{n/2}(V^2+\epsilon^2)^{N-nL/2}}\,,
\label{eq:integrala}
\end{equation}
and if $\epsilon$ and $\delta$ are finite, the numerical integration
can be performed 
using a suitable numerical computation library.

We use $\epsilon$ determined by a (scaled) geometric sequence 
\begin{equation}
\epsilon = \epsilon_l = \epsilon_0/(A_c)^l,\quad(l=0,1,\cdots) 
\label{eq:dcmparm}
\end{equation}
for constants $\epsilon_0, A_c\ (A_c>1)$. Then we expect
\begin{equation}
I=\lim_{l \rightarrow \infty}I(\epsilon_l).
\label{eq:epslim}
\end{equation}

Repeating the numerical integration, we obtain a sequence of
numerical values of $I(\epsilon_l)$ for $l=0,1,\ldots, l_{max}$.
From these values and using an extrapolation method, we can
estimate the value of $I$ with enough accuracy.

Next we discuss the singularity originating from $\delta \rightarrow 0$.
In \siki{integrala}, $V^2+\epsilon^2$ is positive and $U$ is
positive semi-definite. Only at the boundary of the integration
region $U$ becomes 0\ \footnote{
$U$ is a sum of monomials of $x$.}.
When $\delta\rightarrow +0$, it is either integrable
like $\displaystyle{\int_0 dx dy \frac{1}{(x+y)^{1-\delta}}}$ or
non-integrable like $\displaystyle{\int_0 dx \frac{1}{x^{1-\delta}}}$.
In the latter case it develops a pole term $\sim 1/\delta$ as
the ultraviolet singularity\footnote{
The singular pole also appears in the first factor of \siki{integral}
if $N-nL/2\le 0$ for $\delta=0$.}.
Depending on the masses and external momenta, $V$ can be 0
inside the integration region to develop the imaginary part of 
$I$\,~\footnote{
For illustration, one assumes that $N-nL/2=1$ and the variables are
transformed into $V$ and $x'_r$ variables. Then, omitting the 
Jacobian and other details, the imaginary part becomes
$\displaystyle{\int\prod dx'_r %
\int_{V_1}^{V_2} dV \frac{\epsilon}{V^2+\epsilon^2}}$.
In the limit as $\epsilon\rightarrow +0$,
the inner integral is finite if $V_2>0>V_1$ and 0 otherwise.},
and also can be 0 at the boundary of integral region (as in
the case of $U$) to develop an infrared singularity pole $\sim 1/\delta$.
If \siki{integrala} is free from these singularities, we just put $\delta=0$.
If not, the integral in \siki{integrala} is denoted as $I(\delta, \epsilon)$
and we first calculate
$I(\delta)=\lim_{l\rightarrow\infty}I(\delta, \epsilon_l)$
as in \siki{epslim} fixing $\delta$.
Then, we assume the following form:
\begin{equation}
I(\delta)=\cdots + \frac{C_{-1}}{\delta}+C_0 + C_1 \delta+ \cdots
\end{equation}
We calculate $I(\delta)$ for a set of values of $\delta$ and 
estimate the coefficients $C_j$.  For instance, in case of
a single pole, $\delta I(\delta)= C_{-1}+C_0 \delta + O(\delta^2)$ and
we extract $C_{-1}$ and $C_0$\ \footnote{
And also $C_1$ is necessary if the first factor of \siki{integral}
is singular.}.

So much for the description of DCM and one can understand
the necessity of an efficient library for the numerical integration and
that for the extrapolation.
For the former we use {\tt DQAGE}\cite{DQAGE} which is a
variant of Gaussian quadrature. Since it works adaptively,
one can specify the accuracy of the numerical results, although
the high accuracy costs in computation time.
For the latter we use Wynn's  $\epsilon$ algorithm\cite{Wynn}
\footnote{This '$\epsilon$' has nothing to do with the
parameter in propagators.}
which predicts the limiting value by the following iteration.
We set the results of the numerical integration as initial values of 
the series $a(l,k)$:
\begin{equation}
a(l,-1)=0, \qquad a(l,0)=I(\epsilon_l), \qquad l=0,1,\cdots.
\end{equation}
The element $a(l,k+1)$ is obtained 
by the following recurrence relation:
\begin{equation}
a(l,k+1)=a(l+1,k-1)+\frac{1}{a(l+1,k)-a(l,k)}, \qquad l=0,1,\cdots.
\end{equation}
Whilst the $a(l,k)$ values with odd $k$ are meant to store temporary numbers, 
the $a(l,k)$ with even $k$ give extrapolated estimates.

\section{Numerical results}
\label{sec:numerical}

We calculate the integrals for the two-loop box diagrams 
shown in \zu{twoloopb}.
The parameters are $m_1=m_2=m_5=m_6=m=50\mathrm{GeV}$,
$m_3=m_4=m_7=M=90\mathrm{GeV}$,
$p_1^2=p_2^2=p_3^3=p_4^2=m^2$ and $t=(p_1+p_3)^2=-(100)^2\mathrm{GeV^2}$.
We take $s=(p_1+p_2)^2$ variable and introduce
a dimensionless variable $f_s=s/m^2$.

\begin{figure}[htb]
\begin{center}
\unitlength 0.1in
\begin{picture}( 54.8000, 10.3500)(  3.7000,-12.6500)
%
\special{pn 8}%
\special{pa 600 600}%
\special{pa 2600 600}%
\special{fp}%
%
\special{pn 8}%
\special{pa 600 1200}%
\special{pa 2600 1200}%
\special{fp}%
%
\special{pn 8}%
\special{pa 1000 600}%
\special{pa 1000 1200}%
\special{fp}%
\special{pa 1600 600}%
\special{pa 1600 1200}%
\special{fp}%
\special{pa 2200 600}%
\special{pa 2200 1200}%
\special{fp}%
%
\special{pn 8}%
\special{pa 600 540}%
\special{pa 800 540}%
\special{fp}%
\special{sh 1}%
\special{pa 800 540}%
\special{pa 734 520}%
\special{pa 748 540}%
\special{pa 734 560}%
\special{pa 800 540}%
\special{fp}%
%
\special{pn 8}%
\special{pa 600 1260}%
\special{pa 800 1260}%
\special{fp}%
\special{sh 1}%
\special{pa 800 1260}%
\special{pa 734 1240}%
\special{pa 748 1260}%
\special{pa 734 1280}%
\special{pa 800 1260}%
\special{fp}%
%
\special{pn 8}%
\special{pa 2600 1260}%
\special{pa 2400 1260}%
\special{fp}%
\special{sh 1}%
\special{pa 2400 1260}%
\special{pa 2468 1280}%
\special{pa 2454 1260}%
\special{pa 2468 1240}%
\special{pa 2400 1260}%
\special{fp}%
%
\special{pn 8}%
\special{pa 2600 540}%
\special{pa 2400 540}%
\special{fp}%
\special{sh 1}%
\special{pa 2400 540}%
\special{pa 2468 560}%
\special{pa 2454 540}%
\special{pa 2468 520}%
\special{pa 2400 540}%
\special{fp}%
\put(4.0000,-6.0000){\makebox(0,0)[lb]{$p_1$}}%
\put(3.7000,-13.6000){\makebox(0,0)[lb]{$p_2$}}%
\put(26.5000,-13.4000){\makebox(0,0)[lb]{$p_4$}}%
\put(26.4000,-6.1000){\makebox(0,0)[lb]{$p_3$}}%
%
\special{pn 8}%
\special{pa 3800 600}%
\special{pa 5800 600}%
\special{fp}%
%
\special{pn 8}%
\special{pa 3800 1200}%
\special{pa 5800 1200}%
\special{fp}%
%
\special{pn 8}%
\special{pa 3800 540}%
\special{pa 4000 540}%
\special{fp}%
\special{sh 1}%
\special{pa 4000 540}%
\special{pa 3934 520}%
\special{pa 3948 540}%
\special{pa 3934 560}%
\special{pa 4000 540}%
\special{fp}%
%
\special{pn 8}%
\special{pa 3800 1260}%
\special{pa 4000 1260}%
\special{fp}%
\special{sh 1}%
\special{pa 4000 1260}%
\special{pa 3934 1240}%
\special{pa 3948 1260}%
\special{pa 3934 1280}%
\special{pa 4000 1260}%
\special{fp}%
%
\special{pn 8}%
\special{pa 5800 1260}%
\special{pa 5600 1260}%
\special{fp}%
\special{sh 1}%
\special{pa 5600 1260}%
\special{pa 5668 1280}%
\special{pa 5654 1260}%
\special{pa 5668 1240}%
\special{pa 5600 1260}%
\special{fp}%
%
\special{pn 8}%
\special{pa 5800 540}%
\special{pa 5600 540}%
\special{fp}%
\special{sh 1}%
\special{pa 5600 540}%
\special{pa 5668 560}%
\special{pa 5654 540}%
\special{pa 5668 520}%
\special{pa 5600 540}%
\special{fp}%
\put(36.0000,-6.0000){\makebox(0,0)[lb]{$p_1$}}%
\put(35.7000,-13.6000){\makebox(0,0)[lb]{$p_2$}}%
\put(58.5000,-13.4000){\makebox(0,0)[lb]{$p_4$}}%
\put(58.4000,-6.1000){\makebox(0,0)[lb]{$p_3$}}%
%
\special{pn 8}%
\special{pa 4200 600}%
\special{pa 4200 1200}%
\special{fp}%
\special{pa 4800 600}%
\special{pa 5400 1200}%
\special{fp}%
\special{pa 5400 600}%
\special{pa 4800 1200}%
\special{fp}%
\put(12.7000,-5.8000){\makebox(0,0)[lb]{1}}%
\put(44.6000,-5.8000){\makebox(0,0)[lb]{1}}%
\put(12.6000,-13.8000){\makebox(0,0)[lb]{2}}%
\put(44.8000,-13.8000){\makebox(0,0)[lb]{2}}%
\put(8.2000,-10.0000){\makebox(0,0)[lb]{3}}%
\put(40.0000,-10.0000){\makebox(0,0)[lb]{3}}%
\put(16.5000,-10.0000){\makebox(0,0)[lb]{4}}%
\put(22.5000,-10.0000){\makebox(0,0)[lb]{7}}%
\put(18.6000,-5.8000){\makebox(0,0)[lb]{5}}%
\put(18.5000,-13.8000){\makebox(0,0)[lb]{6}}%
\put(48.9000,-10.0000){\makebox(0,0)[lb]{4}}%
\put(50.6000,-13.8000){\makebox(0,0)[lb]{6}}%
\put(50.7000,-5.9000){\makebox(0,0)[lb]{5}}%
\put(52.4000,-10.0000){\makebox(0,0)[lb]{7}}%
\put(4.0000,-4.0000){\makebox(0,0)[lb]{(a)}}%
\put(36.0000,-4.0000){\makebox(0,0)[lb]{(b)}}%
\end{picture}%
\end{center}
\caption{(a)Two-loop planar box diagram and (b) 
Two-loop non-planar box diagram.
The mass of internal line $k$ is $m_k$. Each external
momentum flows inward. }
\label{fig:twoloopb}
\end{figure}
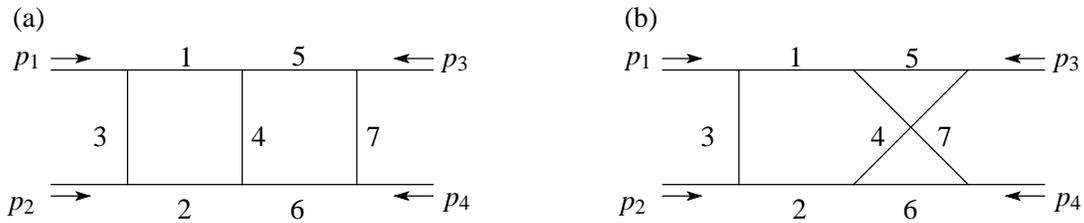

The explicit form of the $U$ and $W$ functions is found in \cite{yuasacpc}.
Since there is no ultraviolet/infrared divergence, 
we put $\delta=0$ in \siki{integral}.
The integral is 6-dimensional and we perform a transformation of
the integration variables onto a 6-dimensional hypercube $[0,1]^6$.
By this transformation one can cancel common variables 
between the numerator and the denominator.
The results are presented in \zu{tbl-fsp} and in \zu{tbc-fsp}.
In \cite{yuasacpc}, we verified the results by
comparison with another computation which is a combination
of algebraic transformations and numerical integration, and also
by the consistency check between the real and imaginary parts through
the dispersion relation.

\iffigureon
\begin{figure}[htb]
\centering
\includegraphics[width=0.6\linewidth]{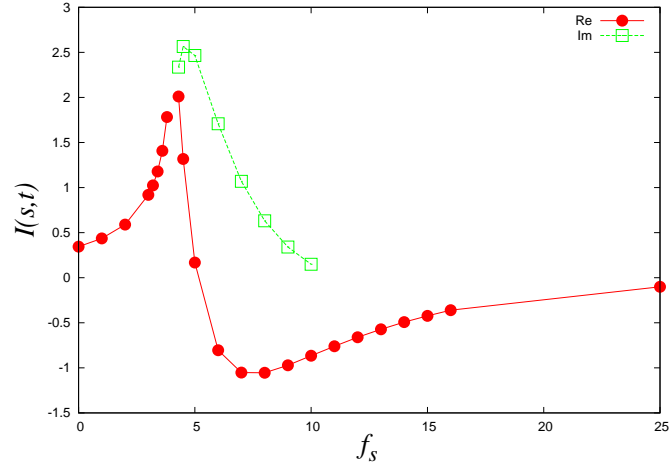}
\caption{Numerical results of $I$ for the planar diagram 
in units of $10^{-12}$ ${\rm GeV}^{-6}$ for $0.0 \le f_{s} \le 25.0$ and 
$t=-10000.0 {\rm GeV}^2$. Plotted points are the real part (bullets) and the imaginary part (squares). For the latter, the region $f_s>10$ is not yet computed.}
\label{fig:tbl-fsp}
\end{figure}
\fi

\iffigureon
\begin{figure}[htb]
\centering
\includegraphics[width=0.6\linewidth]{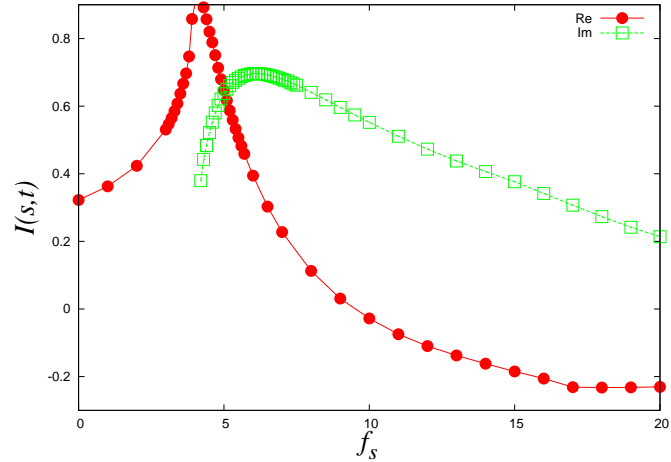}
\caption{Numerical results of $I$ for the non-planar diagram
in units of $10^{-12}$ ${\rm GeV}^{-6}$ for $0.0 \le f_{s}\le 20.0$ and 
$t=-10000.0 {\rm GeV}^2$. Plotted points are the real part (bullets) and the imaginary part (squares).}
\label{fig:tbc-fsp}
\end{figure}
\fi

\iffigureon
\begin{figure}[bht]
\centering
\input{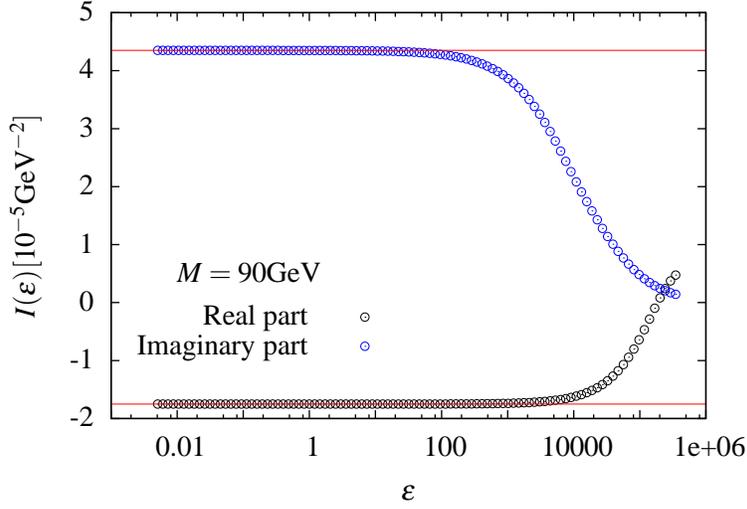}
\caption{The real part(lower points) and imaginary part(upper points)
 of one-loop scalar vertex integral
$I(\epsilon)$ are shown.
The horizontal red line is the exact (analytical) value.}
\label{fig:koike}
\end{figure}
\fi


In DCM, the validity of the extrapolation depends on the
choice of the values of $\epsilon$.
We keep finite value of $\epsilon$ in the numerical integration,
so that its physical dimension is the same as the squared mass.
We have two parameters in \siki{dcmparm}. 
$A_c$ is normally $A_c=2$
and we can use $A_c = 1.2$ or $1.3$  to obtain a less 
computational intensive sequence of integrals as $\epsilon$ 
decreases more slowly.
In order to study the choice of $\epsilon_0$, the initial
value of the iteration, we have calculated the following
on-mass-shell one-loop vertex integral as an example:
\begin{equation}
I= \int_{0\le x+y\le 1} dx dy 
\frac{1}{M^2(1-x-y)+m_e^2(x+y)^2-sxy-i\epsilon}
\end{equation}
Here, $s=500^2\mathrm{GeV^2}$ and $m_e=0.5\times 10^{-3} \mathrm{GeV}$.
The value of this integral is computed with a given value of $M$
and $\epsilon=1.2^{70-m}$ for $m=0,1, \ldots,120$.
Then we use the values for $m, m+1,\ldots, m+14$ as the
input of extrapolation. This means that we take $l_{max}=14$ and 
$\epsilon_0=1.2^{70-m} \mathrm{GeV^2}$ for $m=0,1, \ldots, 100$.

\vspace{3mm}

\noindent
{\bf Table.1}\  Extrapolated values of $I$. 15 values,
$I(\epsilon_l)=\epsilon_0/(1.2)^l, \ (l=0,\ldots,14)$, are
used for the extrapolation. Error is not the difference
from the analytical value but estimated from the
extrapolation.
{
\footnotesize 
\renewcommand{\arraystretch}{0.8}
\begin{center}
\begin{tabular}{rrrrr}
\hline
$\epsilon_0$ & Real part of $I$ & error & Imaginary part of $I$ & error 
\\
\hline
3.49E+05 & -1.75104242540072E-05 & 1.65E-09 
& 2.25556360020931E-09 & 1.80E-09 \\
5.63E+04 & -1.75105247961553E-05 & 3.02E-13 
& -2.96789310051170E-05	& 8.78E-08 \\
9.10E+03 & -1.75105248407057E-05 & 1.55E-14 
& 4.35402513918612E-05 & 1.82E-09 \\
1.47E+03 & -1.75105250993412E-05 & 6.78E-21 
& 4.34982757936633E-05 & 4.04E-13 \\
2.37E+02 & -1.75105250996915E-05 & 2.76E-16 
& 4.34982757936633E-05 & 4.04E-13 \\
3.83E+01 & -1.75105250989527E-05 & 1.40E-17 
& 4.34982788205288E-05 & 1.92E-17 \\
6.19E+00 & -1.75105250987944E-05 & 3.58E-18 
& 4.34982788206820E-05 & 5.55E-16 \\
1.00E+00 & -1.75105251104117E-05 & 1.22E-15 
& 4.34982788654878E-05 & 9.93E-14 \\
1.62E-01 & -1.75105250921691E-05 & 1.98E-16 
& 4.34982793676347E-05 & 2.11E-14 \\
2.61E-02 & -1.75105251352866E-05 & 1.71E-15 
& 4.34982790401757E-05 & 4.97E-14 \\
4.21E-03 & -1.75105251110498E-05 & 4.07E-15 
& 4.34982788582295E-05 & 2.21E-17 \\
\hline
(analytical) & -1.75105250974494E-05 & & 4.34982788194091E-05  & \\
\hline
\end{tabular}
\end{center}
}

The calculated results for $M=90\mathrm{GeV}$ are shown in \zu{koike} and
Table.1.
It is to be noted that $\epsilon$ in
DCM is obviously finite.
One can see that even when the value of $I(\epsilon)$ differs from the
analytical value, the extrapolation gives a good estimation.
There is a finite region that shows the agreement
between the analytical value and the extrapolated one.
This behavior demonstrates
that DCM is stable up to some extent for the choice of
$\epsilon_0$ parameters.
We have tested the similar analysis for several values of $M$
($M=1, 10^1, 10^2, 10^3 \mathrm{GeV}$)
and found similar behavior.
Though we need more tests for this point,
we can temporally conclude that $\epsilon_0$ is not need
to be very small but it can be set to
the typical squared mass in the denominator of the integrand.

In the calculation of the two-loop box diagrams described above,
we have checked the convergence of the extrapolation step-by-step.
The value used for $\epsilon_0$ is $1.2^{40}\sim 1.2^{45} \mathrm{GeV^2}$
which would be consistent with the above conjecture.

\section{Summary}
\label{sec:summary}

In this paper, we have outlined  DCM and
calculated the two-loop scalar box integral as an example to
show its applicability.
Since the radiative correction in the electroweak theory (or in the SUSY
model) involves various combinations of mass parameters
in the integrand,
DCM is a good candidate to
handle general loop integrals.

In order to use DCM for the calculation 
of higher-order radiative corrections, 
we plan to proceed to the following research.

\begin{enumerate} \itemsep=0pt
\item 
The method should be tested for a wider class of
diagrams with various combinations of masses and external
momenta and with the numerator structure.

\item 
Further study on the choice of parameters $\epsilon_0, A_c$
is required for a stable application.

\item 
The variable transformation is sometimes important 
for good convergence.  This is to be processed
in an automatic manner.

\item 
In dimensional regularization, the ultraviolet/infrared 
divergence appears as a pole $1/\delta$.
The separation of the infrared pole is already done successfully
in \cite{infra}.
A similar treatment of ultraviolet poles is expected.

\item 
Sometimes DCM needs long computational time.
It will be important to perform the computations in 
a parallel computing environment.

\end{enumerate}

{\small
\noindent{\bf{Acknowledgements}}

We wish to thank Prof. T.Kaneko for his discussions and valuable comments. 
This work was supported in part 
by the Grant-in-Aid (No.20340063 and No.23540328) of JSPS and 
by the CPIS program of Sokendai.
}


\addtolength{\baselineskip}{1pt}

\end{document}